\documentclass{obspm}
\usepackage{graphicx}
\begin{document}

\TitreGlobal{Hunt for Molecules}

\title{Dust and Molecules in Early Galaxies: 
Prediction and Strategy for Observations}
\author{Tsutomu T.\ Takeuchi}\address{
Laboratoire d'Astrophysique de Marseille (LAM), FRANCE.}
\runningtitle{Dust and Molecules in Early Galaxies}
\setcounter{page}{23}
\index{Takeuchi, T.\ T.}
%
\begin{abstract} 
The interplay between dust and molecules is of fundamental importance
in early galaxy evolution.
First we present the prediction for the dust emission from forming galaxies. 
Then we discuss the observational strategy for molecules in early galaxies 
by infrared absorption lines of a bright continuum source behind the clouds.
By combining these two approaches, we will be able to have a coherent 
picture of the very early stage of galaxy evolution.
\end{abstract}
\maketitle
%
\section{Introduction}

At the very first phase of the Universe, there have been almost no structures, 
as seen in the tiny fluctuation of the cosmic microwave background.
In contrast, today we observe a variety of cosmic structures, from planes to
the large-scale structure in the Universe.
In addition, the Big Bang nucleosynthesis produced only light elements, while
the present Universe is rich in heavy elements.
Thus, galaxy formation may be the most important event in the cosmic history.

Active star formation is followed by heavy element production from the birth
and death of stars.
The produced heavy elements generally exist in the form of small solid 
particles, i.e., dust.
Dust grains play very important roles in galaxy evolution.
First, they scatter and absorb short wavelength photons and finally
re-emit the energy in the infrared (IR).
This makes the galaxy evolutions seen in the optical/ultraviolet and IR
very different with each other (see Takeuchi et al. 2005c).
Another important role of dust grains is that they work as a catalyst of
molecular formation (e.g., Hirashita et al.\ 2002, Hirashita \& Ferrara 2002).
This process is particularly important in the very early phase of galaxy 
evolution, because hydrogen molecules are very important coolant of gas
in the very metal-poor phase of galaxies.
Without dust, star formation does not proceed effectively 
(Hirashita \& Ferrara 2002).

{}From this point of view, we focus on the two kinds of systems.
We first discuss young galaxies with active dust production. 
We will see that their appearance is different from dusty IR luminous 
galaxies at lower redshift (e.g., Takeuchi et al., 2001a,b).
We discuss the observability of their continuum radiation from dust.
Then we move the topic to dense gas systems with little metal/dust.
On the eve of the active dust production, they might contain a large amount
of hydrogen molecules. 
We consider the direct measurement of the hydrogen molecules in such 
protogalactic clouds through the absorption lines in the IR.

\section{Dust Emission from Forming Galaxies}

\subsection{SED Model for Forming Galaxies}\label{sec:model}

We present a brief outline of our model framework. 
All the details are shown in Takeuchi et al.\ (2003, 2005b),
and Takeuchi \& Ishii (2004).

Nozawa et al.\ (2003) investigated the formation of dust grains in 
the ejecta of Population III SNe, whose progenitors are initially metal-free. 
They considered unmixed and uniformly mixed cases in the He core.
In the unmixed case, the original onion-like structure of elements is 
preserved, and in the mixed case, all the elements are uniformly mixed
in the helium core.

For the chemical evolution, we use a closed-box model with
the Salpeter initial mass function $\phi(m) \propto m^{-2.35}$
with mass range of $(m_{\rm l}, m_{\rm u})= (0.1\;M_\odot,100\;M_\odot)$.
We neglect the contribution of SNe Ia and winds from low-mass 
evolved stars to the formation of dust, because we consider the timescale 
younger than $10^9\;\mbox{yr}$.
The interstellar medium is treated as one zone, and the growth of 
dust grains by accretion is neglected.
Within the short timescale considered here, it can be assumed safely.
We also neglect the destruction of dust grains within the young age
considered.
Finally, we assumed a constant SFR for simplicity.

Very small grains cannot establish thermal equilibrium with the ambient 
radiation field, which is called stochastic heating.
For the specific heat, we adopt a multidimensional Debye model 
(e.g., Draine \& Li 2001) for carbon and silicate grains.
For other species, we adopt the classical three-dimensional Debye model
with a single Debye temperature.
The emission from dust is calculated basically according to Draine \& 
Anderson (1985).
Total dust emission is obtained as a superposition of the emission from 
each grain species.
We constructed $Q(a, \lambda)$ of each grain species from available 
experimental data via Mie theory.
When the dust opacity becomes large, we should consider the self-absorption
of IR emission by dust.
This is treated by a thin shell approximation.

\subsection{Results and Discussions}\label{sec:results}

\subsubsection{Evolution of IR SED}

\begin{figure}[h]  
\begin{center}
\centering\includegraphics[angle=90,width=5.7cm]{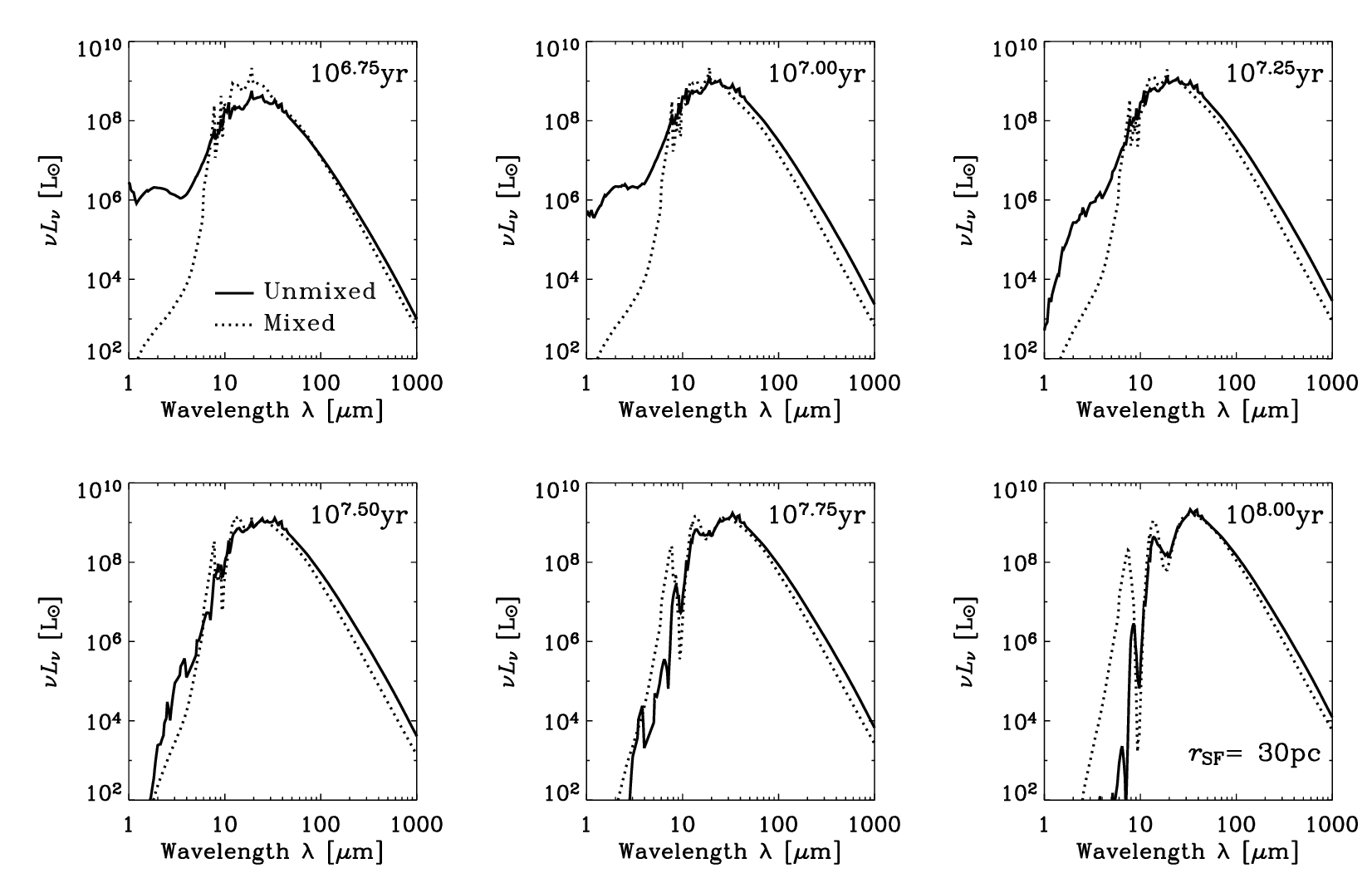}
\centering\includegraphics[angle=90,width=5.7cm]{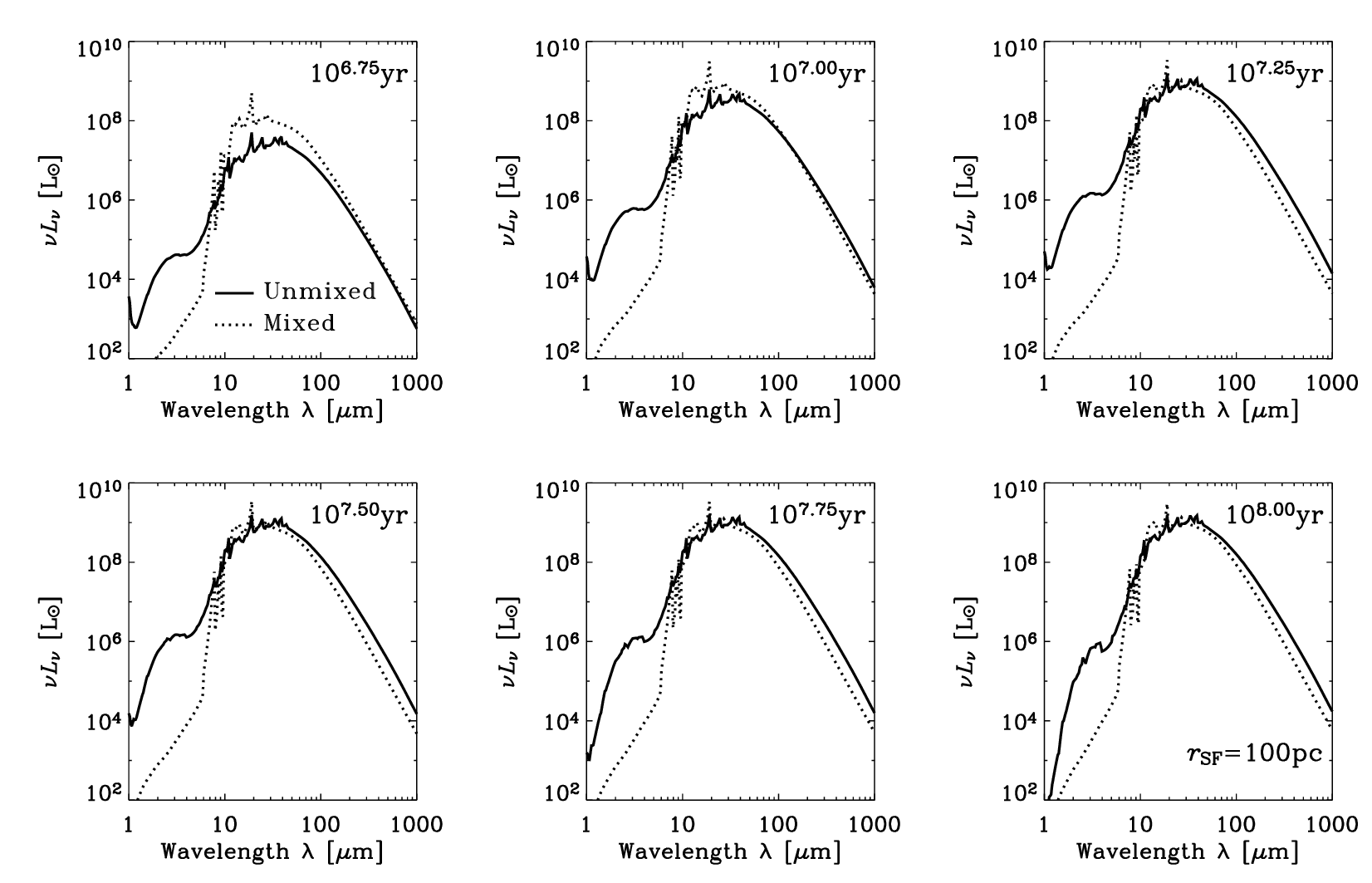}
\end{center}
\caption{
The evolution of the IR SED of a very young galaxy
(left six panels: $r_{\rm SF}=30\,\mbox{pc}$;
right six panels: $100\,\mbox{pc}$).
}\label{fig:sed}
\end{figure} 
\noindent
We first show the evolution of the IR SED of forming galaxies based on 
our baseline model in Figure~\ref{fig:sed}.
For these calculation we adopted the star formation rate ${\rm SFR}=1\,M_\odot
\,\mbox{yr}^{-1}$.
We adopt $r_{\rm SF} = 30$~pc and 100~pc.
In a very young phase ($\mbox{age}=10^{6.75}\mbox{--}10^{7.25}$~yr), 
unmixed-case SED has an enhanced N--MIR continuum.
After $10^{7.25}$~yr, the N--MIR continuum is extinguished by the 
self-absorption in the case of $r_{\rm SF}=30$~pc. 
In contrast, the self-absorption is not significant for $r_{\rm SF}=100$~pc.
In both cases, the SEDs have their peaks at a wavelength $\lambda \simeq 
20\mbox{--}30\;\mu$m.

\subsubsection{A nearby forming galaxy SBS~0335$-$052}

\begin{figure}[h]  
\begin{center}
\centering\includegraphics[width=5cm]{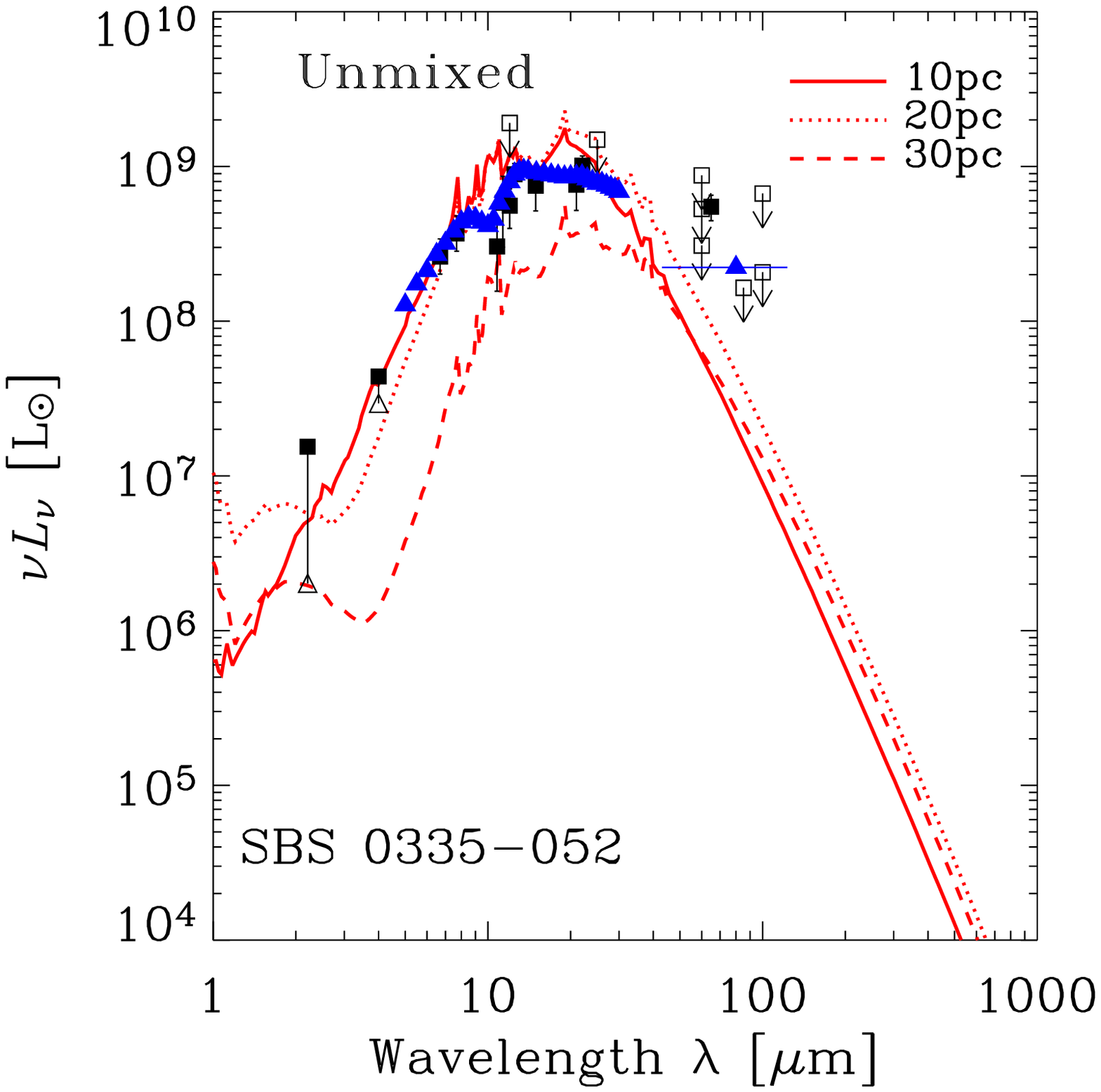}
\centering\includegraphics[width=5cm]{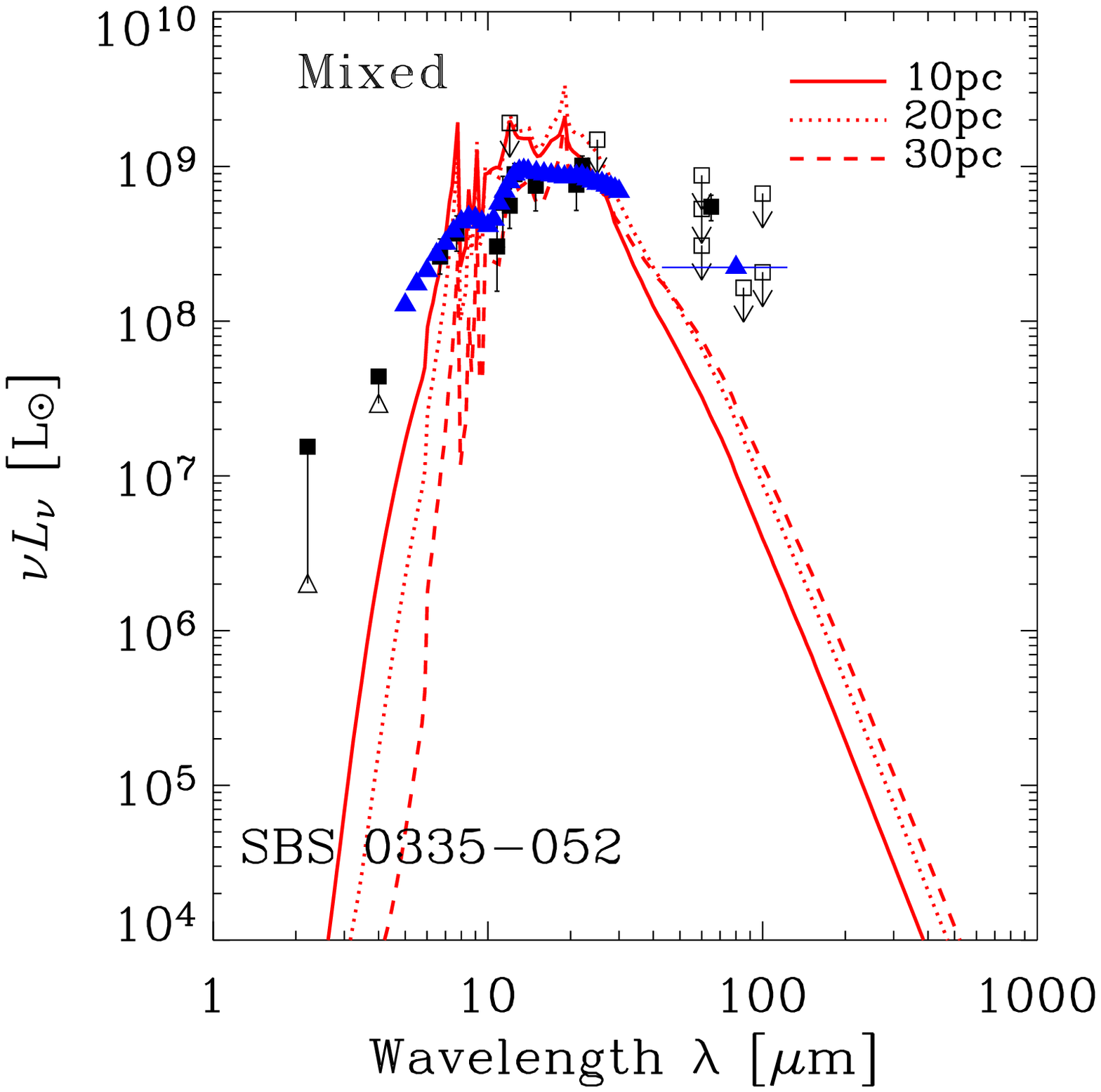}
\end{center}
\vspace*{-0.25cm}  
\caption{The model for the SED of a nearby star-forming dwarf galaxy 
SBS~0335$-$052.
}\label{fig:sed_sbs}
\end{figure} 

\noindent
Although it is still difficult to observe galaxies in their very first phase 
of the SF, we can compare the model with a local counterpart of such
galaxies.
SBS~0335$-$052 is a local galaxy ($\sim 54\;\mbox{Mpc}$) with 
$\mbox{SFR} = 1.7 \;M_\odot\,\mbox{yr}^{-1}$ and extremely 
low metallicity $Z = 1/41\,Z_\odot$.
This galaxy is known to have an unusual IR SED and strong flux at N--MIR.
It has a very young starburst ($\mbox{age} < 5\,\mbox{Myr}$) without 
significant underlying old stellar population.
Houck et al.\ (2004) presented new data of the MIR SED by {\sl Spitzer}.
We have calculated the SED for $r_{\rm SF}=10$, 20, and 30~pc.
The SFR is fixed to be $1.7\,M_\odot\,\mbox{yr}^{-1}$, and the age is 
$10^{6.5}$~yr.
We present the model and the observed SED of SBS~0335$-$052 in 
Figure~\ref{fig:sed_sbs}.
The very strong N--MIR continuum of SBS~0335$-$052 is well reproduced by 
the SED of unmixed case, but the mixed case seriously underpredicts its
MIR continuum.
This suggests that we may determine the dust production (unmixed or
mixed) of SNe through the observation of the N--MIR SEDs of forming galaxies.

\subsubsection{Toward higher redshifts}

\begin{figure}[h]
\begin{center}
\centering\includegraphics[angle=90,width=10cm]{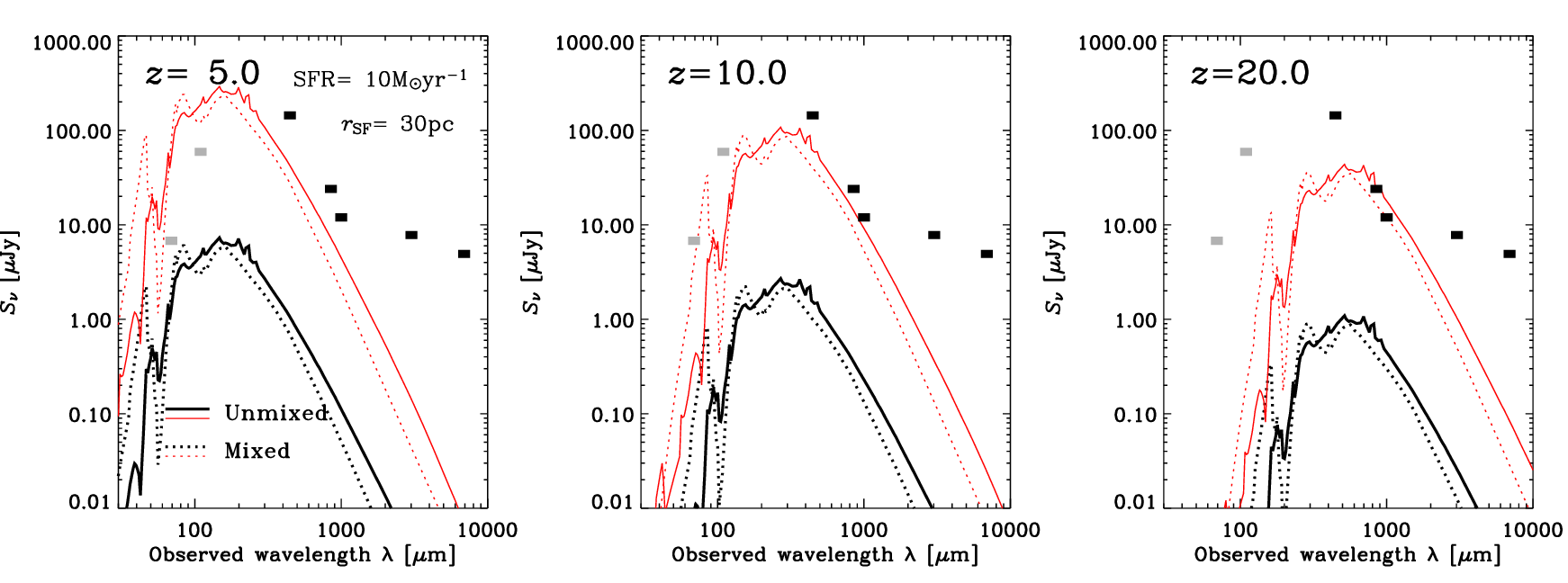}
\centering\includegraphics[angle=90,width=10cm]{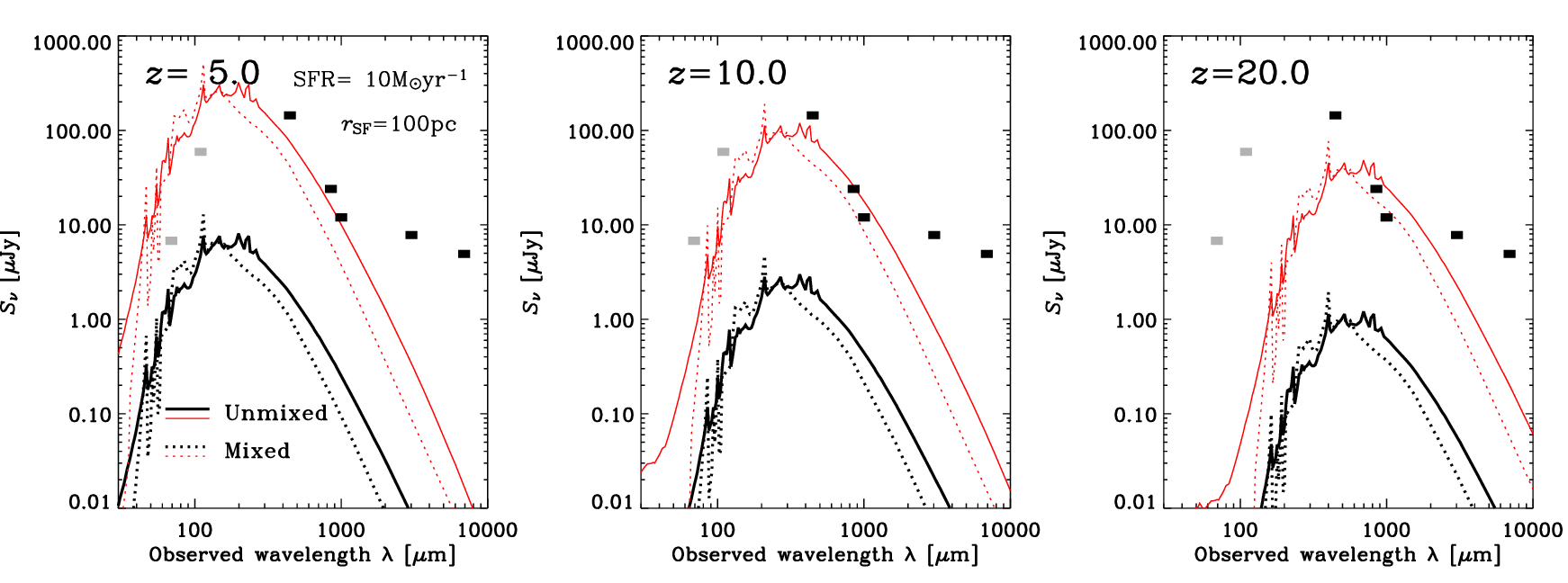}
\end{center}
\caption{The expected flux densities of a dwarf forming galaxy at
$z =5, 10$, and 20. 
Gray and black tick marks are the confusion limits of {\sl Herschel}
and the detection limits of an ALMA 8-hour survey.
The {\sl Herschel} confusion limits are taken from Lagache et al.\ (2003).
}\label{fig:dwarf_highz}
\end{figure}
\noindent
Consider a dark halo of mass $\sim 10^9\;M_\odot$, then it is expected to 
contain a gas with mass $\simeq 10^8\;M_\odot$.
We calculate the SEDs for a dwarf forming galaxy.
If gas collapses on the free-fall timescale with an efficiency of 
$\epsilon_{\rm SF}$ (we assume $\epsilon_{\rm SF}=0.1$), we obtain
the following evaluation of the SFR (Hirashita \& Hunt 2004):
\begin{eqnarray}
  \mbox{SFR} \simeq 0.1 \left(\frac{\epsilon_{\rm SF}}{0.1}\right)
    \left(\frac{M_{\rm gas}}{10^7\;M_\odot}\right)^{3/2}
    \left(\frac{r_{\rm SF}}{100\;\mbox{pc}}\right)^{-3/2} \;
    [M_\odot \mbox{yr}^{-1}]\;.
\end{eqnarray}
If we consider $M_{\rm gas}\simeq 10^8\;M_\odot$, we have $\mbox{SFR}\simeq
3(r_{\rm SF}/100\;\mbox{pc})^{-3/2}\;M_\odot\,\mbox{yr}^{-1}$.
Thus, we consider a dwarf galaxy with $\mbox{SFR} = 10\;M_\odot\mbox{yr}^{-1}$
as an example, and we adopt $r_{\rm SF} = 30$~pc and 100~pc.
The age is set to be $10^7$~yr.
We show the expected SEDs for such galaxies at $z=5$, 10, and 20 in 
Figure~\ref{fig:dwarf_highz}.
It seems almost impossible to detect such objects by {\sl Herschel} or ALMA.
However, gravitational lensing works very well as a natural huge telescope.
This is depicted by the thin lines in Figure~\ref{fig:dwarf_highz}
(with magnification factor 40).
If we suppose a cluster of galaxies at $z_{\rm l}\simeq 0.1\mbox{--}0.2$ 
whose dynamical mass $M_{\rm dyn}$ is $5 \times 10^{14}\,M_\odot$, and
use number counts of high-$z$ forming galaxies presented by Hirashita \& 
Ferrara (2002), we have an expected number of galaxies suffering a strong 
lensing to be $\simeq 1\mbox{--}5$ with detection limit $1\;\mu$Jy.

\section{Direct Measurement of H$_2$ by IR Absorption Lines}

\subsection{Basic Idea}

Molecular hydrogen is the predominant constituent of the dense gas in the 
Universe. 
While in the local Universe, molecules containing heavy elements (e.g.\
CO, ${\rm H_2O}$) are good tracers of the amount of molecular hydrogen, 
they are expected to be significantly depleted in the early Universe.
We need a technique for measuring molecular hydrogen directly. 
Petitjean et al.\ (2000) have reported a direct measurement of ${\rm H}_2$
molecules in the UV.
The transition probabilities of ionization and dissociation lines are 
so large that they are useful for detecting thin layers and 
small amounts of the molecular gas, but useless for detecting dense gas clouds.

In contrast, molecular hydrogen has well-known vibrational and 
rotational transitions at IR wavelengths.
Their transition probabilities are very small because hydrogen molecule, 
a diatomic molecule of two identical nuclei, has no allowed dipole 
transitions.
Hence, they are useful tools to analyze dense ($> 10\; {\rm cm}^{-3}$) and 
hot ($> 300$~K) gas. 
Ciardi \& Ferrara (2001) calculated expected emission line intensities of 
molecular hydrogen, but unfortunately, their direct measurement was
found to be very difficult due to their weakness. 
However, if there is a strong IR continuum source behind or 
in the molecular gas cloud, absorption measurements of these lines may be 
possible.
This idea is presented by Shibai et al.\ (2001).

We should note that the absorption by dust in the cloud is 
usually larger than the absorption by ${\rm H}_2$ molecules. 
However, since our target is a primordial gas cloud whose metallicity is 
significantly lower than the Galactic value, 
the lines can be detected in absorption against bright IR sources.
Such observation will be feasible with the advent of proposed space missions 
for large IR telescope facilities.
Here we focus on {\sl SPICA} (Ueno et al.\ 2000).

\subsection{Calculation}

Assuming a uniform, cool gas cloud ($kT_{\rm ex} \ll h\nu$), the optical 
thickness of the line absorption, $\tau_{\rm line}$ is 
\begin{eqnarray}
\tau_{\rm line} \simeq \frac{\lambda^3}{8\pi}
\left(\frac{g_u}{g_\ell}\right) A_{u\ell}N_\ell \frac{1}{\Delta V}
\end{eqnarray}
where the subscripts $u$ and $\ell$ indicate the upper and lower levels of a 
transition, $g_{u}$ and $g_{\ell}$ are the degeneracy of each 
state respectively, $A_{u\ell}$ is the Einstein's $A$~coefficient, 
$N_{\ell}$ is the column density of the molecules in the lower state, 
and $\Delta V$ is the line width in units of velocity. 
Here we assume that almost all the molecules occupy the lowest energy state.
The absorption line flux in the extinction free case, $I^{\rm abs}_{\rm 
line,0}$, is obtained by
\begin{equation}
  I_{\rm line,0}^{\rm abs}=S\Delta\nu
  \left(1-e^{-\tau_{\rm line}}\right)
\end{equation}
where $\Delta \nu$ is the line width in units of frequency and $S$ is 
the continuum flux density of the IR source behind the cloud.
Assumed parameters in this calculation are listed in Table~\ref{tab2}.
Next, we consider about the dust extinction which is denoted by
\begin{equation}\label{eq:tau_dust}
  \tau_{\rm dust}(\lambda )= 1.086 \left( \frac{A_\lambda}{A_V}\right) 
  \left( \frac{A_V}{N_{\rm H}}\right)_\odot  Z \, N_{\rm H}
\end{equation}
The extinction spectrum of Mathis (1990) is adopted for 
($A_{\lambda}$/$A_V$), and the extinction efficiency is assumed to be 
proportional to the relative heavy element abundance, 
and $(A_V/N_{\rm H})_{\odot}$ is the conversion factor from $A_V$ to 
$N_{\rm H}$ for the local abundance.
We then obtain the absorption line flux with extinction, 
$I^{\rm abs}_{\rm line}$, 
\begin{eqnarray}
  I_{\rm line}^{\rm abs} = 
    I_{\rm line}^{\rm abs,0} e^{-\tau_{\rm dust}}
  = S\Delta\nu \left(1-e^{-\tau_{\rm line}}\right) e^{-\tau_{\rm dust}} \;.
\end{eqnarray}

\begin{table*}[t]
\vspace*{-1cm}
\begin{center}
\caption{Parameters assumed for the present calculation}
\label{tab2}
\vspace*{2mm}
\begin{tabular}{cc} \hline \hline
Ortho : Para & 3 : 1 \\
HD/${\rm H}_2$ & $10^{-5}$ \\
Intrinsic Flux of Source & 10~mJy \\
Line width & 100 ${\rm km\,s}^{-1}$ \\
Detection Limit ({\sl SPICA} 5$\sigma$) & 
$5\times 10^{-21}\; {\rm W\,m}^{-2}$\\
Detectable Optical Thickness & $> 0.01$ \\
Dust Extinction Model & Mathis (1990) \\
\hline
\end{tabular}
\end{center}
\vspace*{-5mm}
\end{table*} 

\subsection{Results}

\begin{figure}[h]
\centering\includegraphics[width=5cm]{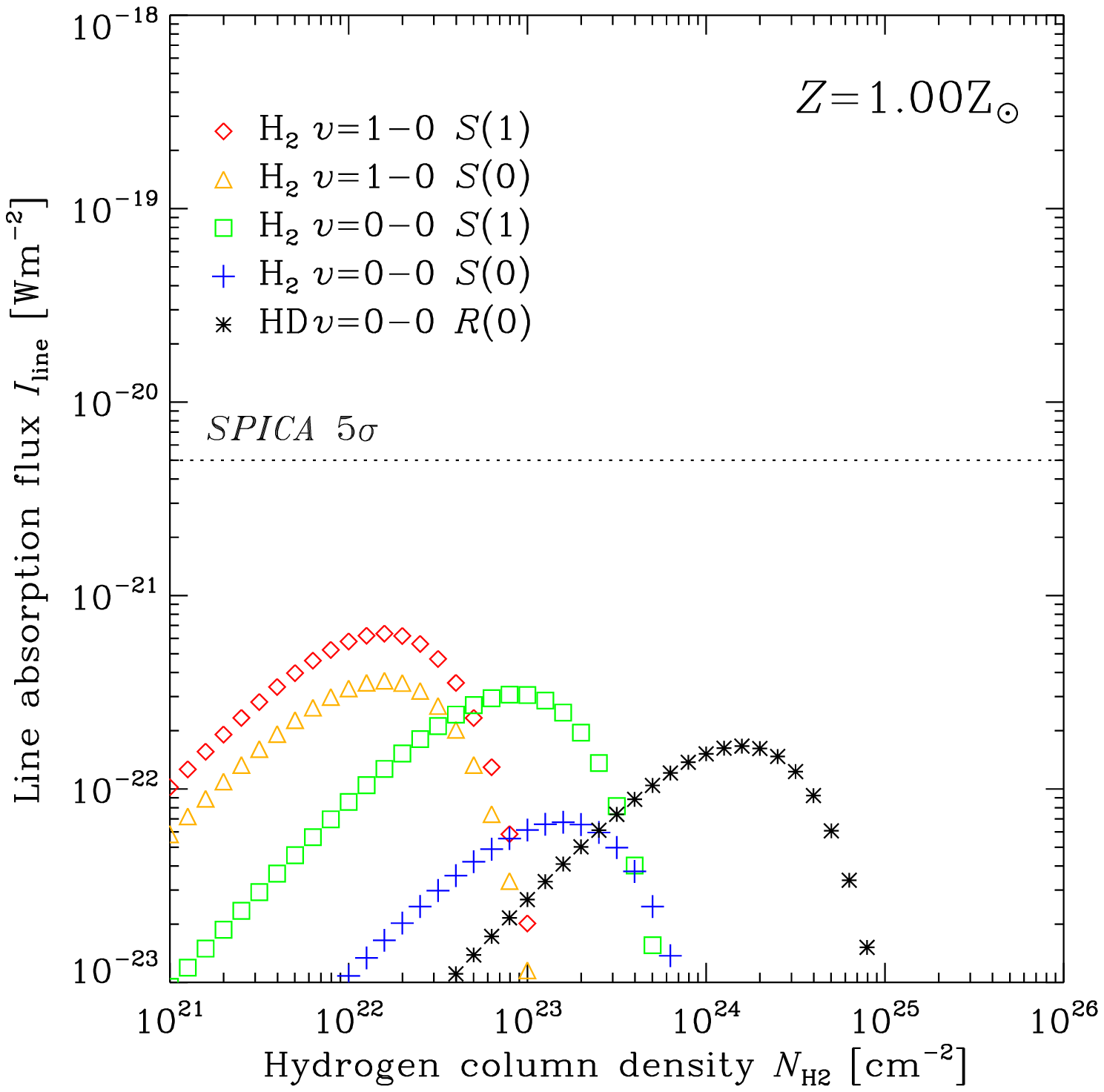}
\centering\includegraphics[width=5cm]{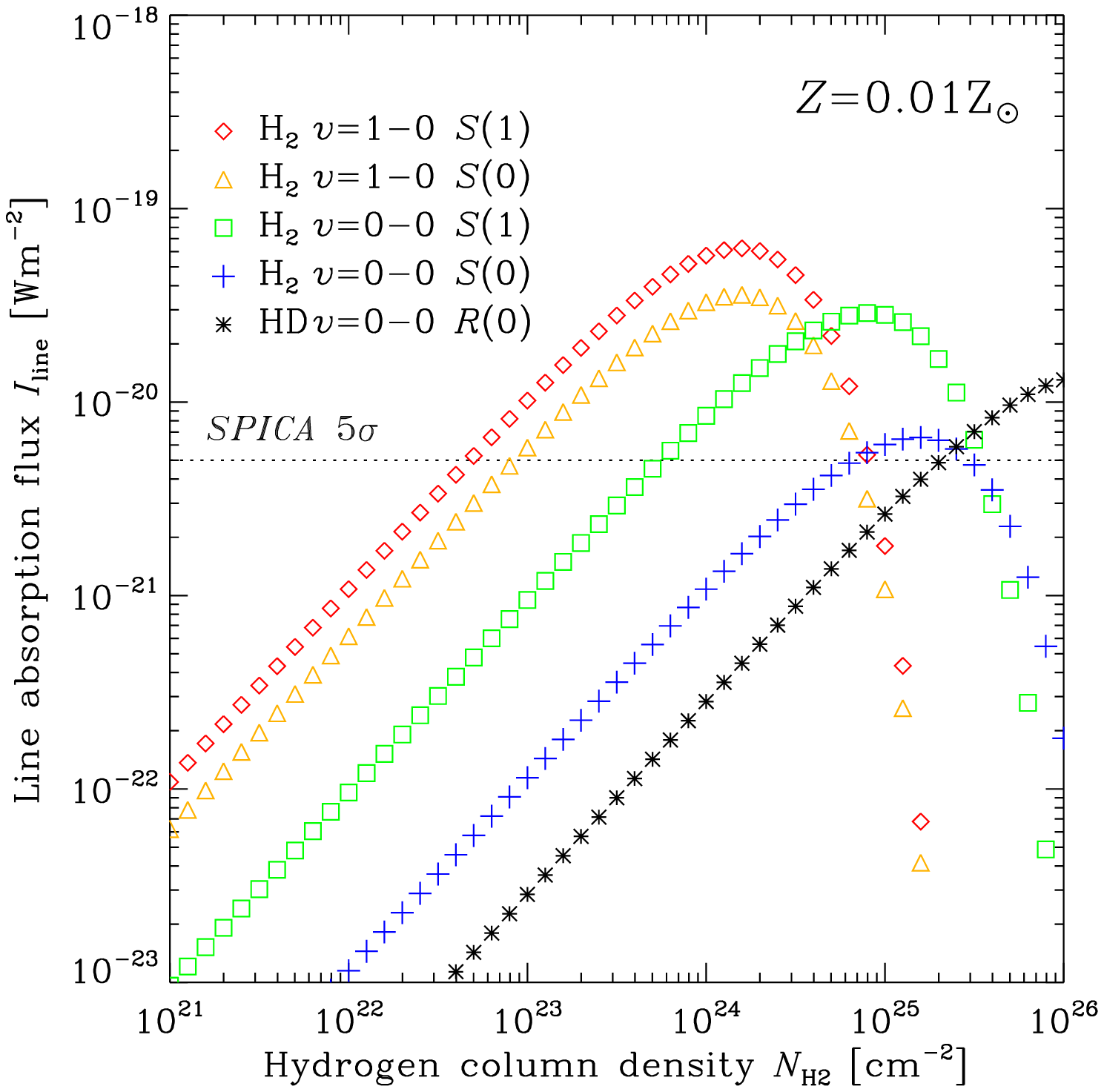}
\caption{
  Left panel: the absorption line flux expected for a cloud with the Milky Way 
  heavy element abundance ($Z = 1$) in front of a 10~mJy source.
  Right panel: same as $(a)$ but the heavy element abundance is $Z = 0.01$.
}\label{fig:abs}
\end{figure}

\begin{table*}[t]
\begin{center}
\caption{Parameters of the Line Transitions}\label{tab3}
\vspace*{2mm}
\begin{tabular}{ccllc} \hline \hline
Transition & Wavelength & $A$ coefficient & $A_{\lambda}/A_V$\\
~& [$\mu$m]  & [${\rm s}^{-1}$] & \\
\hline
${\rm H_2}\; v = 1-0$ $S(1)$ & 2.12 & $3.47 \times 10^{-7}$ & 0.11\\
${\rm H_2}\; v = 1-0$ $S(0)$ & 2.22 & $2.53 \times 10^{-7}$ & 0.11\\
${\rm H_2}\; v = 0-0$ $S(1)$ & 17 & $4.77 \times 10^{-10}$ & 0.020\\
${\rm H_2}\; v = 0-0$ $S(0)$ & 28 & $2.95 \times 10^{-11}$ & 0.011\\
HD $v = 0-0$ $R(0)$ & 112 & $2.54 \times 10^{-8}$ & 0.0011\\
\hline
\end{tabular}
\end{center}
\vspace*{-7mm}
\end{table*} 

\noindent
The calculation was made for the five lines listed in Table~\ref{tab3}. 
Left panel of Figure~\ref{fig:abs} shows the absorption line flux expected 
for a cloud with the local heavy element abundance in front of a 10~mJy 
source. 
The absorption fluxes are far smaller than the $5 \sigma $ expected 
sensitivity of the {\sl SPICA} mission (Ueno et al.\ 2000). 
On the other hand, right panel of Figure~\ref{fig:abs} shows the result 
for the case in which the heavy element abundance is 1~\% of that of the 
local one. 
All five lines populate parameter space above or near the limit.

We have assumed an intrinsic continuum flux density of 10~mJy at the line 
wavelength.
We examined some existing QSOs as template sources (for the details, see 
Shibai et al.\ 2001).
If we have objects at $z = 5$ at least as luminous as our template 
sources, it is possible to observe absorption in ${\rm H_2}$ lines of 17, 28,
and 112~$\mu$m respectively.
The redshifted 112~$\mu$m line will be in the submm range and may not be 
observable by IR missions, but ALMA should be able to observe it.
Unfortunately, it is very difficult to observe the absorption of 2.2~$\mu$m
lines since the expected flux density based on dust emission spectrum 
will be too faint.

\subsection{What can we learn from the absorption lines of these clouds?}

First we simply estimate the column density of a protogalactic 
hydrogen cloud.
Here we assume that the size of the cloud $R$ is a few kpc 
and that the gas mass $M$ is $\sim 10^{11}\;M_\odot$.
Such a large reservoir of molecular gas has been discovered at high redshift 
(Papadopoulos et al.\ 2001). 
Such clouds have column density of
\begin{eqnarray}
  N_{{\rm H}_2} \simeq 4 \times 10^{23} [\mbox{cm}^{-2}]\, f
    \left(\frac{R}{3\;\mbox{kpc}} \right)^{-2}
    \left( \frac{M}{10^{11}M_\odot}\right)
\end{eqnarray}
where $f$ is the mass fraction of the molecular clouds to the total gas mass,
and $M$ is the total gas mass of the protogalaxy.
The radius will be finally a few kpc.
In a realistic situation, a primordial cloud may evolve dynamically on
a free-fall timescale, which is much shorter than the timescales of 
cosmological structure evolution.
Therefore, observed properties are specific to the redshift at which the
cloud absorption is measured.
Hence, since we will obtain redshift $z$, and velocity dispersion $\Delta V$, 
we can trace the dynamical evolution of a primordial cloud at high-$z$.
The {\sl SPICA} mission will provide us with the information on the massive
objects ($> 10^{11}~M_\odot$) at $z< 5$. 
We still have to wait for more sensitive facilities for the observational
approach to the Population III objects, whose physical properties 
are theoretically predicted by recent extensive investigations, 
because they are located at higher redshift and their typical mass is 
small (e.g., Nishi \& Susa 1999).
When we estimate the detectability of such objects by other facilities, 
the algorithm constructed in this paper can be applied straightforwardly.

\section{Conclusion}\label{sec:conclusion}

We first present our dust emission model from forming galaxies,
which is based on a new theory of SN II dust production.
The model roughly reproduced the observed SED of a local low-metallicity
dwarf SBS~0335$-$052 which has a peculiar strong and MIR-bright dust SED.
We also calculated the SED of a very high-$z$ forming small galaxy.
Although it may be intrinsically too faint to be detect ed even by ALMA
8-hour survey, the gravitational lensing can make it feasible.

Then we proposed a method to measure the amount of H$_2$ in primordial
low-metallicity cloud in an IR spectra of QSOs.
If metallicity of the cloud is low ($Z \sim 0.01 Z_\odot$), dust extinction
is so weak that 17 and $28\;\mu$m lines are detectable by {\sl SPICA} for
objects at $z < 5$.
For very high-$z$ Population III objects, ALMA will be useful.
By this observation, we can trace back the dynamical evolution of early 
collapsing objects at very high-$z$.

By combining these two approaches, we will be able to have a coherent 
picture of the very early stage of galaxy evolution, from the pristine
gas phase to active dust production phase.

\bigskip
This work is made through collaborations with many people, especially
Andrea Ferrara, Hiroyuki Hirashita, Leslie K.\ Hunt, Takako T.\ Ishii, 
Takashi Kozasa, Takaya Nozawa, and Hiroshi Shibai (alphabetic order).
I have been supported by the Japan Society of the Promotion of Science
as a Postdoctoral Fellow for Research Abroad (Apr.\ 2004--Dec.\ 2005).

\end{document}